\documentclass[submitting]{nst}
\usepackage{subfigure,dcolumn}
\usepackage[T2A,T1]{fontenc}
\usepackage[russian,english]{babel}

\usepackage{listings}
\begin{document}

\title{A scheme for excitation of thorium-229 nuclei based on the electronic bridge excitaion}\thanks{Supported by the National Natural Science Foundation of China (Grant No. 11804372)}

\author{Lin Li}
\affiliation{State Key Laboratory of Magnetic Resonance and Atomic and Molecular Physics, Innovation Academy for Precision Measurement Science and Technology, Chinese Academy of Sciences, Wuhan430071, China}
\affiliation{University of Chinese Academy of Sciences, Beijing 100049, China}
\author{Zi Li}
\affiliation{State Key Laboratory of Magnetic Resonance and Atomic and Molecular Physics, Innovation Academy for Precision Measurement Science and Technology, Chinese Academy of Sciences, Wuhan430071, China}
\affiliation{University of Chinese Academy of Sciences, Beijing 100049, China}
\author{Chen Wang}
\affiliation{State Key Laboratory of Magnetic Resonance and Atomic and Molecular Physics, Innovation Academy for Precision Measurement Science and Technology, Chinese Academy of Sciences, Wuhan430071, China}
\affiliation{University of Chinese Academy of Sciences, Beijing 100049, China}
\author{Wen-Ting Gan}
\affiliation{State Key Laboratory of Magnetic Resonance and Atomic and Molecular Physics, Innovation Academy for Precision Measurement Science and Technology, Chinese Academy of Sciences, Wuhan430071, China}
\affiliation{University of Chinese Academy of Sciences, Beijing 100049, China}
\author{Xia Hua}
\email[Xia Hua, ]{huaxia@wipm.ac.cn}
\affiliation{State Key Laboratory of Magnetic Resonance and Atomic and Molecular Physics, Innovation Academy for Precision Measurement Science and Technology, Chinese Academy of Sciences, Wuhan430071, China}
\author{Xin Tong}
\email[Xin Tong, ]{tongxin@wipm.ac.cn}
\affiliation{State Key Laboratory of Magnetic Resonance and Atomic and Molecular Physics, Innovation Academy for Precision Measurement Science and Technology, Chinese Academy of Sciences, Wuhan430071, China}

\begin{abstract}
Thorium-229 possesses the lowest nuclear first excited state with an energy of about 8 eV. The extremely narrow linewidth of the nuclear first excited state with the uncertainty of 53 THz prevents the direct laser excitation and the realization of the nuclear clock. We present a proposal using the Coulomb crystal of a linear chain formed by the $^{229}$Th$^{3+}$ ions, the nuclei of $^{229}$Th$^{3+}$ ions in the ion trap are excited by the electronic bridge (EB) process. The 7$P_{1/2}$ state of the thorium-229 nuclear ground state is chosen for the EB excitation. Using the two-level optical Bloch equation under experimental conditions, we calculate that 2 out of 36 prepared thorium ions in the Coulomb crystal can be excited to the nuclear first excited state, and it takes about 2 hours to scan over the uncertainty of 0.22 eV. Taking the advantage of transition enhancement of the EB and the long stability of the Coulomb crystal, the energy uncertainty of the first excited state can be limited to the order of 1 GHz.
\end{abstract}
\maketitle

\section{Introduction}
\label{intro}
The excitation energy between the nuclear ground state and the nuclear first excited state (isomeric state) of thorium-229 (Th-229) is about 8 eV \cite{1,2,3,4}. The optical transition linewidth of the nuclear first excited state is $10^{-4}$ Hz \cite{5,6}. The low isomeric energy and narrow linewidth provide potential for a resonator quality with the factor of $10^{19}$ and make Th-229 the most suitable choice for the development of nuclear optical clocks \cite{6,7,8}. Such a Th-229 nuclear clock is expected to be a sensitive probe for time variation of the fundamental constants of nature; besides, it will open opportunities for highly sensitive tests of fundamental principles of physics, particularly in searches for violations of Einstein’s equivalence principle and new particles \cite{9,10,11}. \par
The isomeric energy is currently measured in two ways: 1) Measuring the kinetic energy of the internal conversion electrons from the Th-229 isomeric state. For example, an electron spectrometer was used to detect the internal conversion electron energy and the isomeric energy obtained was 8.28 $\pm$ 0.17 eV \cite{4}. 2) Observing the $\gamma$ radiation from the Th-229 excited nuclei. An isomeric energy of 7.8 $\pm$ 0.5 eV was obtained by a magnetic microcalorimeter \cite{2}. Later, a 29.2 keV inter-band excitation of the Th-229 nuclear state was observed using synchrotron radiation \cite{3}. Combining with the in-band 29.2 keV transition observed via nuclear rotational spectroscopy, the Th-229 isomeric state energy was determined to be 8.30 $\pm$ 0.92 eV \cite{12}. Recently, a more precise magnetic microcalorimeter with improved energy resolution was developed, the isomeric energy of 8.10 $\pm$ 0.17 eV was obtained \cite{1}. Taking different weights for these measurements, the average value of the isomeric state is determined to be 8.12 $\pm$ 0.11 eV \cite{13}. Its uncertainty corresponds to 53 THz, which is at least 16 orders of magnitude higher than the narrow linewidth of $10^{-4}$ Hz.\par

\begin{figure}[hb]
\centering
\setlength{\belowcaptionskip}{0cm}
\includegraphics[width=7cm]{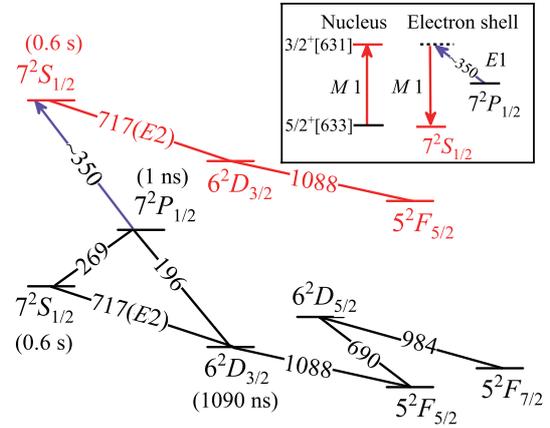}
\caption{Diagram of electronic energy levels and electric-dipole transitions of Th-229 triply charged ions. Inserts show the processes of the EB excitation, the red arrows are the 150 nm $M$1 transition. The black horizontal lines indicate the nuclear ground state of Th-229 and the red horizontal lines indicate the isomeric state of Th-229, respectively. The purple arrow is the 350 nm EB excitation. The lifetime of electronic excited state is indicated in the parentheses. Optical transition wavelengths are in nm and the integers near atomic levels indicate the principal quantum numbers.}
\label{fig:1}
\end{figure}

To improve the precision of the isomeric energy, the synchrotron radiation is used to irradiate the Th-229 doped in a vacuum ultraviolet (VUV) transparent crystal, and then the $\gamma$ photons are detected from the decay of the Th-229 excited nuclei, this measurement excludes roughly half of the favored transition search area \cite{14}. Another scheme is proposed to achieve the Th-229 nuclei excitation by electron capture, and detect the isomeric state of ions in heavy-ion storage rings \cite{15}. Recently, an optical frequency comb is proposed to irradiate the Th-229 dioxide film, the isomeric energy can be measured based on the internal conversion process \cite{16}. Besides, a new approach is suggested to excite the isomeric Th-229 nuclear state via a laser-driven electron recollision. The advantage of the approach is that it does not require knowledge of the isomeric energy precisely \cite{17}.

In this paper, we propose to excite the nuclei of $^{229}$Th$^{3+}$ ions in the ion trap by the EB excitation and compare with the direct nuclear excitation by a pulsed laser. As shown in the insets of Fig.~\ref{fig:1}, the EB excitation starts from an electric dipole ($E$1) transition that is utilized to excite the electron from an initial state to a virtual intermediate state. Then the intermediate state decays via the magnetic dipole ($M$1) transition, and the nuclear ground state 5/2$^{+}$[633] is excited to the isomeric state 3/2$^{+}$[631] simultaneously. So far, the EB schemes for the nuclear excitation have been theoretically investigated for $^{229g}$Th$^{+}$ \cite{18}, $^{229}$Th$^{2+}$ \cite{19}, $^{229}$Th$^{3+}$ \cite{20,21,22} and $^{229}$Th$^{35+}$ \cite{23}. Our proposed experiment is based on the EB excitation in the triply charged $^{229}$Th$^{3+}$ ions. Due to the simple electronic energy levels, the trapped $^{229}$Th$^{3+}$ ions can be laser-cooled to form a linear-chain Coulomb crystal using two closed transitions in the nuclear ground state of $^{229}$Th$^{3+}$ ($^{229g}$Th$^{3+}$) \cite{24}. When the $^{229g}$Th$^{3+}$ ions are prepared in the state 7$P_{1/2}$, the EB excitation is triggered by a 350 nm laser and isomeric states are expected to be populated. Finally, the isomeric state is detected based on the isomer shift by the cooling lasers \cite{7}.\par

\section{Preparation of $^{229g}$$\rm\textbf{Th}$$^{3+}$ ion Coulomb crystals and the state 7P$_{1/2}$}
The $^{229g}$Th$^{3+}$ ions are typically produced by the laser ablation \cite{24,25}. The linear Paul trap for the trapping  $^{229g}$Th$^{3+}$ ions has been described in detail in our previous works \cite{26,27}. The trapped $^{229g}$Th$^{3+}$ ions are laser-cooled to form a Coulomb crystal.\par

\begin{figure}[!htpb]
\centering
\setlength{\abovecaptionskip}{0cm}
\includegraphics[width=7cm]{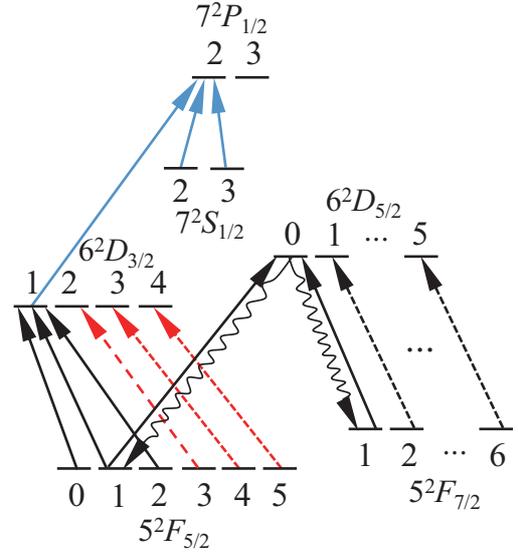}
\caption{\label{fig:0} The state preparation scheme of $^{229g}$Th$^{3+}$ ion. The cooling scheme utilizes the two closed-system indicated by the black solid arrows. The wavy arrows indicate spontaneous radiation and the blue solid arrows indicate the transition to the state 7$P_{1/2}$. The red dashed arrows are the transitions to transfer ions at $\mid5F_{5/2},F=3, 4, 5\rangle$ to $\mid5F_{5/2},F=1\rangle$ and $\mid5F_{5/2},F=2\rangle$. The black dashed arrows are the transitions to transfer ions at $\mid5F_{7/2},F=2, 3, 4, 5,6\rangle$ to $\mid5F_{7/2},F=1\rangle$. The integers near atomic levels indicate total angular momentum quantum numbers $F$.
}
\end{figure}

The energy levels of laser-cooling and fluorescence detecting of the trapped ions are shown in Fig.~\ref{fig:0}. The $^{229g}$Th$^{3+}$ ions are Doppler laser-cooled on $5^{2}F_{5/2}$ $\leftrightarrow$ $6^{2}D_{3/2}$ transition at 1088 nm. Alternatively, a closed optical transition of the three-level $\Lambda$ system, $5^{2}F_{5/2}$ $\leftrightarrow$ $6^{2}D_{5/2}$ $\leftrightarrow$ $5^{2}F_{7/2}$ can also be used to laser-cool the $^{229g}$Th$^{3+}$ ions with two lasers at 690 nm and 984 nm \cite{7}. The population in three hyperfine levels of $\mid5F_{5/2},F=3, 4, 5\rangle$ can be transferred to $\mid5F_{5/2},F=1\rangle$ and $\mid5F_{5/2},F=2\rangle$ via $6^{2}D_{3/2}$ state, and the population in the $\mid5F_{7/2},F=1\rangle$ state is cumulated as shown in Fig.~\ref{fig:0}. Therefore, the ions at other hyperfine levels are transferred to the closed system of $5^{2}F_{5/2}$ $\leftrightarrow$ $6^{2}D_{5/2}$ $\leftrightarrow$ $5^{2}F_{7/2}$ and the second laser cooling cycle is realized. 

\begin{figure}[htpb]
\centering
\setlength{\abovecaptionskip}{0.cm}
\includegraphics[width=8cm]{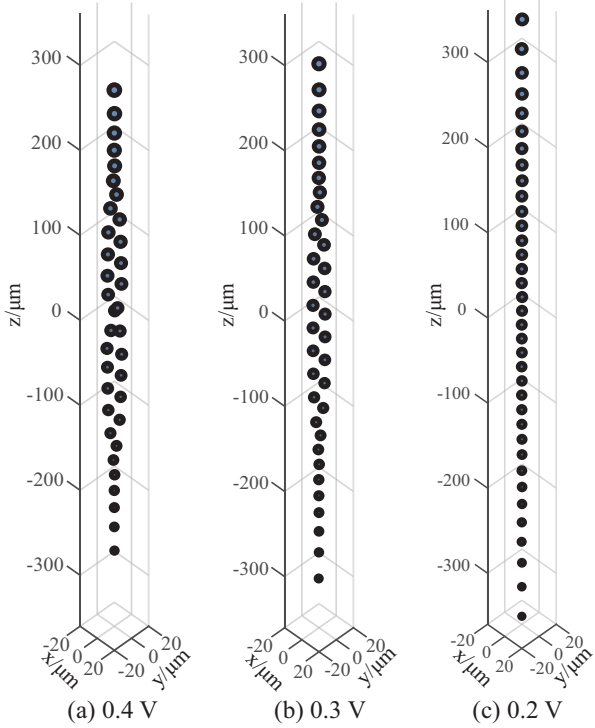}
\caption{\label{fig:3}The simulation results of Coulomb crystal consisting of 36 $^{229g}$Th$^{3+}$ ions. With the same RF voltages (200 $V_{0-pk}$), the Coulomb crystals' shapes are changed under different endcap voltages (a) 0.4 V, (b) 0.3 V and (c) 0.2 V. See text for details.
}
\end{figure}

Once the $^{229g}$Th$^{3+}$ ions are laser-cooled down to 50 $\mu$K, the Doppler broadening of the transition $6^{2}D_{5/2}$ $\rightarrow$ $5^{2}F_{7/2}$ at 984 nm is about 1 MHz \cite{24}. At this temperature, the corresponding Doppler broadenings of the direct nuclear excitation at 150 nm and the EB excitation at 350 nm are estimated to be 6.5 MHz and 2.8 MHz, respectively. The shapes of Coulomb crystals can be manipulated into a linear chain by adjusting the radiofrequency (rf) and endcap voltages \cite{28}. The two-dimensional and linear-chain Coulomb crystals can be simulated with a single ion resolution \cite{29,30}. Using the GPU-accelerated LAMMPS \cite{31} wrapped by the LION package \cite{32}, a Coulomb crystal consisting of 36 $^{229g}$Th$^{3+}$ ions are simulated as shown in Fig.~\ref{fig:3}. When the endcap voltage of 0.2 V and a rf voltage (200 $V_{0-pk}$) of 2 MHz are applied to the ion trap, a linear-chain Coulomb crystal of 36 $^{229g}$Th$^{3+}$ ions (Fig.~\ref{fig:3}(c)) is obtained. When a rf voltage (200 $V_{0-pk}$) of 2 MHz and an endcap voltage of 0.025 V are applied to the ion trap, a linear-chain Coulomb crystal of 100 $^{229g}$Th$^{3+}$ ions is obtained. From the simulation results, the motion amplitude of ions in x- and y-direction (the radial direction) is known to be less than 0.6 $\mu$m when the temperature is below 5 mK, which is consistent with the motion amplitude of single ions \cite{33}.  Therefore, a strongly focused laser spot size of over 1.2 $\mu$m is required for increasing the EB excitation rate and the direct nuclear excitation rate.\par

To excite EB excitation from the state 7$P_{1/2}$ of the $^{229g}$Th$^{3+}$ ions, the three-level transition $5^{2}F_{5/2}$ $\rightarrow$ $6^{2}D_{3/2}$ $\rightarrow$ $7^{2}P_{1/2}$ can be used to accumulate the population at the 7$P_{1/2}$ state. Once the $^{229g}$Th$^{3+}$ ions are laser-cooled down to 50 $\mu$K, the 690 nm and 984 nm laser are turned off in sequence to maintain the ions all at the $5^{2}F_{5/2}$ state, the 1088 nm laser is kept on. To transfer a large population to the $7^{2}P_{1/2}$ state with its lifetime of 1 ns, both 269 nm and 196 nm transitions are excited simultaneously as shown in Fig.~\ref{fig:1}. The natural linewidths of 269 nm transition and 196 nm transition are about 2$\pi$ $\cdot$ 32 MHz and 2$\pi$ $\cdot$ 127 MHz, respectively \cite{34}. The continuous wave (CW) lasers at both the wavelengths are commercially available \cite{35}.\par

The spontaneous radiation rate of the hyperfine structure level transitions between two electronic energy levels is \cite{36}
\begin{eqnarray}
\Gamma_{f\rightarrow i}=\left|\left\langle \xi_{f} J_{f} I^{g} F_{f}\|d\| \xi_{i} J_{i} I^{g} F_{i}\right\rangle \right|^2\frac{4 \omega^3}{3 \hbar c^3} \label{aq0},
\end{eqnarray}
where $\xi_{i}$ ($\xi_{f}$) incorporates all other electronic quantum numbers, $J_{i}$ ($J_{f}$) is the total electronic quantum number of initial (final) state, $F_{i}$ ($F_{f}$) is the total quantum number of initial (final) state, $I^{g}$ is the nuclear spin of nuclear ground state, $d$ is electron dipole, $\omega$ is the angular frequency of the electronic transition, ${\hbar}$ is the reduced Planck constant, $c$ is the speed of light in vacuum. $\left|\left\langle  J_{f} I^{g} F_{f}\|d\| J_{i} I^{g} F_{i}\right\rangle \right|^2$ is the transition strength of the hyperfine structure level, which can be expressed as
\begin{eqnarray}
\left|\left\langle \xi_{f} J_{f} I^{g} F_{f}\|d\| \xi_{i} J_{i} I^{g} F_{i}\right\rangle \right|^2\!=\!(2 F_{i}\!+\!1)(2 F_{f}\!+\!1)\nonumber\\ \!\left\{\begin{array}{ccc} \!J_{f} & F_{f}&I^{g}\!\\ \!F_{i}&J_{i} &1\!\end{array}\right\}^2\!\left|\left\langle  J_{f}\|d\| J_{i} \right\rangle \right|^2 \label{aq10},
\end{eqnarray}
where $\left|\left\langle  J_{f}\|d\| J_{i} \right\rangle \right|^2$ is the transition strength of the electronic energy levels. The spontaneous radiation rate of the electronic energy levels is 
\begin{eqnarray}
\Gamma_{spont}=\left|\left\langle \xi_{f} J_{f}\|d\|\xi_{i} J_{i} \right\rangle \right|^2\frac{4 \omega^3}{3 \hbar c^3}\label{aq11}.
\end{eqnarray}
The excitation rate of the hyperfine structure levels between two electronic energy levels is \cite{37}
\begin{eqnarray}
\Gamma_{exc}=\Gamma_{f\rightarrow i}\frac{\pi^2c^2}{\hbar\omega^3}P_{\omega}\frac{2F_f+1}{2F_i+1}\label{aq12},
\end{eqnarray}
where $P_{\omega}$ is the laser spectral intensity. The stimulated emission rate of the hyperfine structure levels between two electronic energy levels is
\begin{eqnarray}
\Gamma_{sti}=\Gamma_{exc}\frac{2F_i+1}{2F_f+1} \label{aq13}.
\end{eqnarray}
With the laser power of 1 mW and 100 $\mu$m laser spot size for 1088 nm, 269 nm and 196 nm lasers, the steady-state condition can be fulfilled for all $5^{2}F_{5/2}$, $7^{2}S_{1/2}$, $6^{2}D_{3/2}$ and $7^{2}P_{1/2}$ states under continuous illumination. For the multiple hyperfine structure transitions between electronic transitions $5^{2}F_{5/2}$ $\leftrightarrow$ $6^{2}D_{3/2}$ and $7^{2}S_{1/2}$ $\leftrightarrow$ $7^{2}P_{1/2}$, the power of the lasers is divided equally. Thus, after state preparation, the population of the $\mid7P_{1/2},F=2\rangle$ state accounts for about 12{\%} of the total number of trapped ions.
\section{The calculation of the population in the nuclear excited state}
The time-dependent nuclear excitation probability $\rho_{exc}(t)$ for a single nucleus under resonant irradiation is given by Torrey’s solution of the optical Bloch equations as the following \cite{38,39,40}: \par

\textcircled{1} $\quad \Omega_{eg}>\frac{|\Gamma-\tilde{\Gamma}|}{2}$,
\begin{eqnarray}
\rho{_{exc}}(t)=&&\frac{\Omega_{eg}^2}{2(\Gamma\tilde{\Gamma}+\Omega_{eg}^2)}[1-e^{-\frac{1}{2}(\Gamma+\tilde{\Gamma}) t}\nonumber\\
&&\times(\cos({\lambda} t)+\frac{\Gamma+\tilde{\Gamma}}{2\lambda}\sin(\lambda t))]\label{aq1},
\end{eqnarray}

\textcircled{2} $\quad \Omega_{eg}<\frac{|\Gamma-\tilde{\Gamma}|}{2}$,
\begin{eqnarray}
\rho{_{exc}}(t)=&&\frac{\Omega_{eg}^2}{2(\Gamma\tilde{\Gamma}+\Omega_{eg}^2)}[1-e^{-\frac{1}{2}(\Gamma+\tilde{\Gamma}) t}\nonumber\\
&& \times(\cosh({\lambda} t)+\frac{\Gamma+\tilde{\Gamma}}{2\lambda}\sinh(\lambda t))]\label{aq2}.
\end{eqnarray}

Here $\Omega_{eg}$ denotes the Rabi frequency between the isomeric state and the nuclear ground state. $\Gamma=\Gamma_{\gamma}+\Gamma_{nr}$ is the total linewidth of the isomeric state, $\Gamma_{\gamma}$ denotes the $\gamma$ decay rate of the isomeric state. Because the ionization energy of $^{229g}$Th$^{3+}$ is 28.6 eV is much greater than the energy of the isomeric state, the internal conversion process is strongly forbidden \cite{41}. The possible non-radiation transition linewidth $\Gamma_{nr}$ is then dominated by the electronic bridge transition for $^{229g}$Th$^{3+}$. $\tilde{\Gamma}=\frac{\Gamma+\Gamma_{L}}{2}+\Gamma_{add}$ denotes the total transition linewidth \cite{40}, $\Gamma_{L}$ denotes the laser linewidth used for the nuclear excitation, $\Gamma_{add}$ is the additional incoherent linewidth such as phonon coupling of ions in a Coulomb crystal. Since $\Gamma_{add}$ is much
smaller than the laser linewidth $\Gamma_{L}$, the influence of incoherent
linewidth is negligible. $t$ denotes the interaction time between the laser and the $^{229g}$Th$^{3+}$ ions. Here $\lambda=|\Omega_{eg}^2-(\Gamma-\tilde{\Gamma})^2/4|$.\par

Under the resonant condition, the Rabi frequency between the isomeric state and the nuclear ground state is \cite{39}\par
\begin{eqnarray}
\Omega_{eg}=\sqrt{\frac{2\pi c^2IC_{eg}^2\Gamma}{\hbar{\omega_{m}^{3}}}}.
\label{aq3}
\end{eqnarray}
Here $I$ is the intensity of the excitation laser, $\omega_{m}$ is the angular frequency of the nuclear transition. The Zeeman splitting cuased by the geomagnetic field between the two outermost lines is below 1 kHz and can be negligible. So the Clebsch-Gordan coefficient of the sub-states transition between the isomeric state and the nuclear ground state C$_{eg}$ is 1 \cite{16}. \par
Assuming the intensities of all the lasers have Gaussian distribution, the Rabi frequency can be modified as the following two situations: 
\begin{itemize}
\item[i)]
If the laser linewidth ($\Gamma_{L}$) is greater than the Doppler broadening linewidth ($\Gamma_{D}$),  only a part of the laser with the matched frequency range can interact with the triply charged thorium ions effectively. The equivalent Doppler broadening linewidth is
\end{itemize}
\begin{eqnarray}
\Gamma_{D}^{'}=\frac{1}{2}\sqrt{\frac{\pi}{\ln2}}\Gamma_{D}\label{aq4},
\end{eqnarray}\par
The effective laser intensity is
\begin{eqnarray}
I_{eff}=I\int_{-\frac{\Gamma_{D}^{'}}{2}}^{\frac{\Gamma_{D}^{'}}{2}}g_{L}(\omega)d\omega\label{aq5}.
\end{eqnarray}\par
The modified Rabi frequency is
\begin{eqnarray}
\Omega_{eg}^{'}=\sqrt{\frac{2\pi c^2I_{eff}C_{eg}^2\Gamma}{\hbar{\omega_{m}^{3}}}}\label{aq6}.
\end{eqnarray}\par
\begin{itemize}
\item[ii)]
If the laser linewidth ($\Gamma_{L}$) is smaller than the Doppler broadening linewidth ($\Gamma_{D}$), only a part of ions interact with the laser light is on the frequency resonant at a time, and the equivalent laser linewidth is
\end{itemize}
\begin{eqnarray}
\Gamma_{L}^{'}=\frac{1}{2}\sqrt{\frac{\pi}{\ln2}}\Gamma_{L}\label{aq7},
\end{eqnarray}\par
The effective Doppler broadening linewidth is
\begin{eqnarray}
\Gamma_{eff}=\Gamma\int_{-\frac{\Gamma_{L}^{'}}{2}}^{\frac{\Gamma_{L}^{'}}{2}}g_{D}(\omega)d\omega\label{aq8}.
\end{eqnarray}\par
The modified Rabi frequency is
\begin{eqnarray}
\Omega_{eg}^{'}=\sqrt{\frac{2\pi c^2IC_{eg}^2\Gamma_{eff}}{\hbar{\omega_{m}^{3}}}}\label{aq9}.
\end{eqnarray}\par

\section{Direct nuclear excitation}
Th-229 isomeric energy is 8.12 $\pm$ 0.11 eV, corresponding to a wavelength of 150.7$-$154.8 nm. Currently, the available light sources such as the VUV pulsed laser \cite{42}, the synchrotron radiation light source \cite{14}, and the 7th harmonic of a Yb-doped fiber laser \cite{16,43,44,45} are suitable candidates for the direct nuclear excitation. Here, based on the resonance-enhanced four wave mixing, the parameter of a tunable and pulsed VUV laser source are a pulse energy of $E_{L}$ = 13.2 $\mu$J around 150 nm, a bandwidth of $\delta\nu_{L}$ = 15 GHz, and a repetition rate of $R_{L}$ = 10 Hz, which are used to calculate the direct nuclear excitation rate\cite{42}. \par

\begin{figure}[htp]
\centering
\includegraphics[width=7cm]{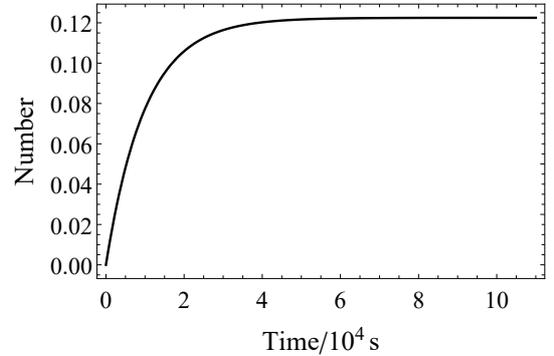}
\caption{\label{fig:4}The expected number of the excited nuclei is calculated as a function of time when 100 triply charged Th-229 ions are directly illuminated by a direct excitation laser.}
\end{figure}

Assuming a strongly focused laser with a 3 $\mu$m spot radius to irradiate the linear chain of 100 ions, so the corresponding laser intensity $I=\frac{E_{L}R_{L}}{\pi r^{2}} = 4.67\cdot10^{6}$ W/$m^{2}$ can be reached. The temperature of Th$^{3+}$ ions is 50 $\mu$K, the corresponding Doppler broadening linewidth $\Gamma_{D}$ = 6.5 MHz is narrower than the laser linewidth. So the modified Rabi frequency on resonance $\Omega_{eg}^{'}(\delta=0)$ = 107 Hz can be obtained based on Eq.~(\ref{aq4})$-$(\ref{aq6}), which is much smaller than the total transition linewidth $\tilde{\Gamma}$ as shown in Table~\ref{tab:1}. In this case, $\Omega_{eg}^{'}<\frac{|\Gamma-\tilde{\Gamma}|}{2}$ is satisfied. The number of excited nuclei is calculated as a function of time based on Eq.~(\ref{aq2}). As shown in Fig.~\ref{fig:4}, the number of excited Th-229 nuclei is only 0.12 out of 100 Th-229 ions and reaches saturated at the irradiation time of 30000 s. Such required laser irradiation time is beyond the stable trapping time, therefore, no Th-229 isomeric state can be detected under the current experimental condition. \par

\begin{table*}[htpb]
\footnotesize
\newcommand{\tabincell}[2]{\begin{tabular}{@{}#1@{}}#2\end{tabular}}
\centering
\caption{\bf Values of variables used for the calculation of the number of excited nuclei based Eq.~(\ref{aq2}).}

\begin{tabular}{cccc}
\hline
\tabincell{c}{Viriable} & 
\tabincell{c}{Description} & 
\tabincell{c}{Value} &
\tabincell{c}{Comment} \\
\hline
$\tilde{\Gamma}$ & Total transition linewidth & 2 $\pi\cdot$ 7.5 GHz & \tabincell{c}{Much larger than the\\ nuclear transition linewidth}\\
$\Gamma_{\gamma}$ & Radiative linewidth & $10^{-4}$ Hz & Estimated from theory\\
$\Gamma_{D}$ & Doppler broadening linewidth & 6.5 MHz & \tabincell{c}{ Ions' Doppler broadenings\\ of the direct nuclear excitation}\\
N & Number of ions & 100 & Laser-cooled ions\\
$\Omega_{eg}$ & Rabi frequency & 1156 Hz & \tabincell{c}{The direct excitation between the isomeric\\  state and the nuclear ground state}\\
$\Omega_{eg}^{'}(\delta=0)$ & Modified Rabi frequency & 107 Hz & Modified Rabi frequency on resonance\\

\hline
\end{tabular}
 \label{tab:1}

\end{table*}
\par

\section{Electronic bridge excitation}
The direct nuclear excitation is limited by the natural linewidth of the isomeric state and the power of the excitation laser. If the EB excitation takes place, the transition linewidth can be increased by about 40 times compared with the direct nuclear excitation \cite{46}. The EB excitation can be utilized by exciting the quantum state of the electronic state $7P_{1/2}$ of $^{229g}$Th$^{3+}$ ($7P_{1/2}$ $5/2^+[633]$) to the electronic state $7S_{1/2}$ of $^{229m}$Th$^{3+}$ ($7S_{1/2}$ $3/2^+[631]$) as shown in Fig.~\ref{fig:1}. The prepared $^{229g}$Th$^{3+}$ Coulomb crystal consisting of a linear chain of 36 ions is shown in Fig.~\ref{fig:3}(c), among which there are 4 ions (12{\%} of the total number of trapped ions) at the state 7$P_{1/2}$ under the steady-state conditions (see Section 2).\par
\begin{table}[htbp]
\footnotesize
\newcommand{\tabincell}[2]{\begin{tabular}{@{}#1@{}}#2\end{tabular}}

\centering
\caption{ Values of variables used for the calculation of the number of excited nuclei based Eq.~(\ref{aq2}).}

\begin{tabular}{ccc}
\hline
\tabincell{c}{Variable} & 
\tabincell{c}{Value} &
\tabincell{c}{Comment} \\
\hline
$\Gamma_{L}$ & 2$\pi\cdot$ 500 kHz & \tabincell{c}{Larger than the \\nuclear transition linewidth}\\
$\Gamma_{\gamma}$ & $10^{-4}$ Hz & Estimated from theory\\
$\Gamma_{D}$ & 2.8 MHz & \tabincell{c}{Ions' Doppler broadenings \\of the EB excitation}\\
N & 36 & Laser-cooled ions\\
$\alpha_{eb}$ & 40 & EB decay for Th$^{3+}$ ions\\
\tabincell{c}{$\Omega_{eg}$} & \tabincell{c}{334 kHz}  & \tabincell{c}{Rabi frequency}\\
$\Omega_{eg}^{'}(\delta=0)$ & 140 kHz  & Modified $\Omega_{eg}$ on resonance\\
$\Omega_{eg}^{'}(\delta=\frac{\Gamma_{D}}{2})$ & 98 kHz & Modified $\Omega_{eg}$ detune $\frac{\Gamma_{D}}{2}$\\
\hline
\end{tabular}
\label{tab:2}

\end{table}
The wavelength of the laser for the EB excitation is 350 nm. A 350 nm laser with the power of 250 mW can be obtained from Toptica DLC TA{-}SHG pro and the laser is focused to a spot size of 10 $\mu$m. Then an intensity of 7.95 $\cdot$ $10^{8}$ Wm$^{-2}$ can be achieved. Therefore, the Rabi frequency is estimated to be 334 kHz.\par

\begin{figure}[htp]
\centering

\includegraphics[width=7cm]{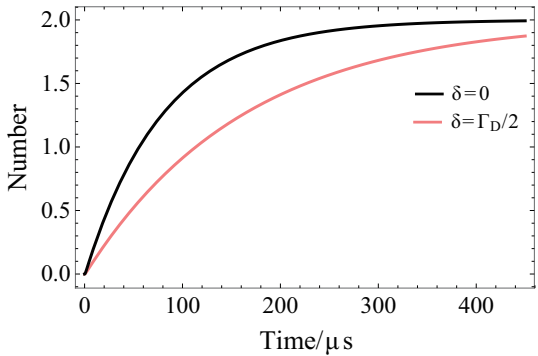}
\caption{\label{fig:5}The expected number of the excited nuclei is calculated as a function of time when 36 triply charged Th-229 ions are trapped and 4 ions are at the state 7$P_{1/2}$.}
\end{figure}
The EB laser linewidth, typically 500 kHz at 350 nm, is narrower than the Doppler broadening of the trapped thorium ions. The modified Rabi frequency $\Omega_{eg}^{'}$ is 140 kHz on resonance and 98 kHz on the detuning $\Gamma_{D}$/2, respectively (shown in Table~\ref{tab:2}).Then the condition $\Omega_{eg}^{'}<\frac{|\Gamma-\tilde{\Gamma}|}{2}$ is satisfied, so the number of excited nuclei as a function of time can be obtained based on Eq.~(\ref{aq2}) and shown in Fig~\ref{fig:5}. If the excitation energy from the state 7$P_{1/2}$ of $^{229g}$Th$^{3+}$ to the state 7$S_{1/2}$ of $^{229m}$Th$^{3+}$ is on resonance, about 2 $^{229g}$Th$^{3+}$ ions can be excited to the isomeric state after 300 $\mu$s irradiation by the 350 nm laser. If the excitation laser is detuned by $\Gamma_{D}/2$, about 1.8 $^{229g}$Th$^{3+}$ ions can be excited to the isomeric state with the same irradiation time. \par

After a successful EB excitation, the Th-229 ions at the isomeric state can be identified based on the different hyperfine structures of the isomeric state and the nuclear ground state. The detailed hyperfine structures of the isomeric state are shown in the Appendix B. After EB nuclear excitation, the EB laser at 350 nm and the state-preparation lasers at 196 nm and 269 nm are both turned off, the cooling lasers at 690 nm and 984 nm are switched back on immediately after and combined with the 1088 nm laser to cool the ions. The nuclear-excited ions stay at the isomeric states and accumulate at the electronic state of $5^{2}F_{5/2}$ as illustrated in Fig.~\ref{fig:1}. The rest nuclear-unexcited ions remain in the nuclear ground state of the Doppler cooling cycle of $5^{2}F_{5/2}$, $6^{2}D_{5/2}$, $5^{2}F_{7/2}$ and $6^{2}D_{3/2}$. Because the isomer shift of $6D_{5/2}$ and $6D_{3/2}$ electronic state are both about 400 MHz \cite{24,47}. The 690 nm and 984 nm lasers are off-resonance for the Doppler cooling of ions at the isomeric state. As a result, the nuclear-excited $^{229m}$Th$^{3+}$ ions leave the 690 nm and 984 nm lasers' cooling cycle and appear as the dark ions in the Coulomb crystal \cite{7}. In the linear-chain $^{229}$Th$^{3+}$ ions Coulomb crystal, single ions are distinguishable and any dark $^{229m}$Th$^{3+}$ ions in the dark generated by a successful EB excitation event can be detected. \par
\begin{table}[!htbp]
\footnotesize
\newcommand{\tabincell}[2]{\begin{tabular}{@{}#1@{}}#2\end{tabular}}
\centering
\caption{ Main parameters for 350 nm irradiation on 36 trapped ions.}
\begin{tabular}{cc}
\hline
\tabincell{c}{Variable} & \tabincell{c}{Value}\\
\hline
Time per scan step & $\approx$ 300 $\mu s$\\
Number of ions on the state 7$P_{1/2}$ & 4\\
Number of laser-cooled ions & $\approx 36$\\
Number of isomeric states per scan step & 2($\delta=0$)\\
Number of isomeric states per scan step & 1.8($\delta=\frac{\Gamma_{D}}{2}$)\\
Time required to scan 0.22 eV & $\approx$ 5670 s\\
\hline
\end{tabular}
 \label{tab:3}
\end{table}
The current reported uncertainty of the isomeric state energy is 0.22 eV \cite{13}. To cover this uncertainty range, it requires about 1.9 $\cdot$ $10^{7}$ scans for each scanning interval of the Doppler broadening linewidth $\Gamma_{D}$. The time for a single scan step takes about 300 $\mu$s, therefore the total scan time takes about 5670 s as shown in Table~\ref{tab:3}. The final isomeric energy is determined by the sum of the energy interval between the electronic state 7$P_{1/2}$ of  $^{229g}$Th$^{3+}$ and the electronic state 7$S_{1/2}$ of $^{229m}$Th$^{3+}$ and the photon energy of the 350 nm laser. Since the uncertainty of the natural linewidth of the state 7$P_{1/2}$ of $^{229g}$Th$^{3+}$ is 1 GHz which is much larger than the laser linewidth and Doppler broadening linewidth, the uncertainty of isomeric energy is on the order of 1 GHz (corresponding to $4 \cdot 10^{-6}$ eV) dominated by the natural linewidth of the state 7$P_{1/2}$.

\section{ Conclusion}
In this paper, taking account of the Doppler broadening of thorium ions in the ion trap, we show that 2 $^{229g}$Th$^{3+}$ ions out of 36 trapped ions can be excited to the isomeric state under the resonant condition via the EB excitation. We present a feasible proposal that a total measurement time of about 2 hours can be achieved for the current uncertainty of the isomeric state energy 0.22 eV. If a cryogenic linear Paul trap is used, the ions can be stably trapped for a longer period of time and the radiative lifetime of the isomeric state can be measured. The utilization of the EB excitation can reduce the uncertainty of Th-229 isomeric energy to about 1 GHz. The calculation is based on the theoretical prediction that the transition linewidth of the EB excitation is 40 times larger than the direct nuclear excitation. If the EB excitation is not observed as proposed in this paper, it may indicate that the coupling between the electrons and the nuclear core is not expectedly strong or the explored nuclear transition energy range is not accurate, and the laser energy used for the EB excitation needs to be increased. On the other hand, successfully observing the EB excitation will further improve the accuracy of isomeric energy, pave the way for the development of a nuclear optical clock, test the temporal variation of fundamental constants and provide new methods for studying nuclear physics.

\appendix
\setcounter{section}{1}
\section*{Appendix A. The experimental steps}

\begin{figure}[!htpb]
\centering
\setlength{\abovecaptionskip}{0.cm}
\includegraphics[width=7cm]{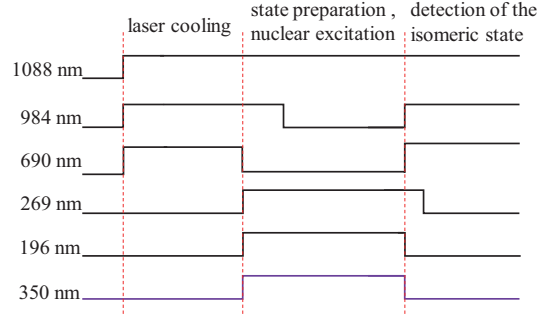}
\caption{\label{AppA1}The experimental steps and the time of each laser switch.}
\end{figure}

 As shown in Fig.~\ref{AppA1}, the experiments are divided into three steps:\par
\textcircled{1} Laser cooling: the 1088 nm, 984 nm and 690 nm lasers are turned on to laser-cool the ions into a Coulomb crystal of a linear chain. Later, the 690 nm laser and 984 nm laser is turned off in sequence to keep all the population at the state $5F_{5/2}$.\par
\textcircled{2} State preparation and nuclear excitation: once the 690 nm laser is turned off, the 269 nm and 196 nm lasers are turned on to prepare the $^{229g}$Th$^{3+}$ ions at the state of 7$P_{1/2}$ and the 350 nm laser is turned on to excite the Th-229 nuclei from the nuclear ground state to the isomeric state. \par
\textcircled{3} Detection of the isomeric state: the 350 nm laser is turned off to prohibit the electronic bridge transition from 7$S_{1/2}$ of the isomeric state and maintain the population of isomeric state. Simultaneously, the 196 nm and 269 nm lasers are turned off in sequence to cumulate the ions at the four lowest electronic states. Then, 690 nm and 984 nm lasers are turned on to detect the isomeric state based on the isomer shift.

\appendix
\setcounter{section}{2}
\section*{Appendix B. The hyperfine structures of $^{229}$$\rm\textbf{Th}^{3+}$}

The hyperfine structures of $^{229g}$Th$^{3+}$ are measured as in ref \cite{48}, where $\mu^{g}=0.360(7)\mu_N$ and $Q^{g}=3.11(16)eb$. Later, the hyperfine structures of  $^{229g}$Th$^{2+}$ and $^{229m}$Th$^{2+}$ are measured in ref \cite{47}, where $\mu^{m}=-0.37(6)\mu_N$ and $Q^{m}=1.74(6)eb$.\par
The magnetic dipole ($A^{m}$) hyperfine coefficients of $^{229m}$Th$^{3+}$ are determined by 
 \begin{eqnarray}
 A^{m}=\frac{A^{g}I^{g}\mu^{m}}{\mu_{g}I^{m}}\label{AppBaq1}.
 \end{eqnarray}\par 
 The electric quadrupole ($B^m$) hyperfine coefficients of the $^{229m}$Th$^{3+}$ are determined by 
  \begin{eqnarray}
 B^{m}=\frac{B^{g}Q^{m}}{Q^{g}}\label{AppBaq2}.
 \end{eqnarray}\par 
The magnetic dipole and electric quadrupole
hyperfine coefficients of $^{229m}$Th$^{3+}$ four lowest levels are calculated as shown in Table.~\ref{AppBtab:1}.
\begin{table}[htbp]
\footnotesize
\setlength{\tabcolsep}{4mm}
\newcommand{\tabincell}[2]{\begin{tabular}{@{}#1@{}}#2\end{tabular}}
\centering
\caption{ The magnetic dipole and electric quadrupole hyperfine coefficients of $^{229m}$Th$^{3+}$ four lowest levels.}
\begin{tabular}{ccc}
\hline
\hline
\tabincell{c}{Valence orbital} & 
\tabincell{c}{$A^{m}$}(MHz) & 
\tabincell{c}{$B^{m}$}(MHz) \\
\hline
$5F_{5/2}$ & -140(20) & 1270(80) \\
$5F_{7/2}$ & -54(9) & 1430(90) \\
$6D_{3/2}$ & -270(40) & 1270(80) \\
$6D_{5/2}$ & 22(4) & 1510(90)\\
\hline
\hline
\end{tabular}
 \label{AppBtab:1}
\end{table} \par
\begin{figure*}[!htpb]
\centering
\setlength{\abovecaptionskip}{0cm}
\setlength{\belowcaptionskip}{0cm}
\includegraphics[width=10cm]{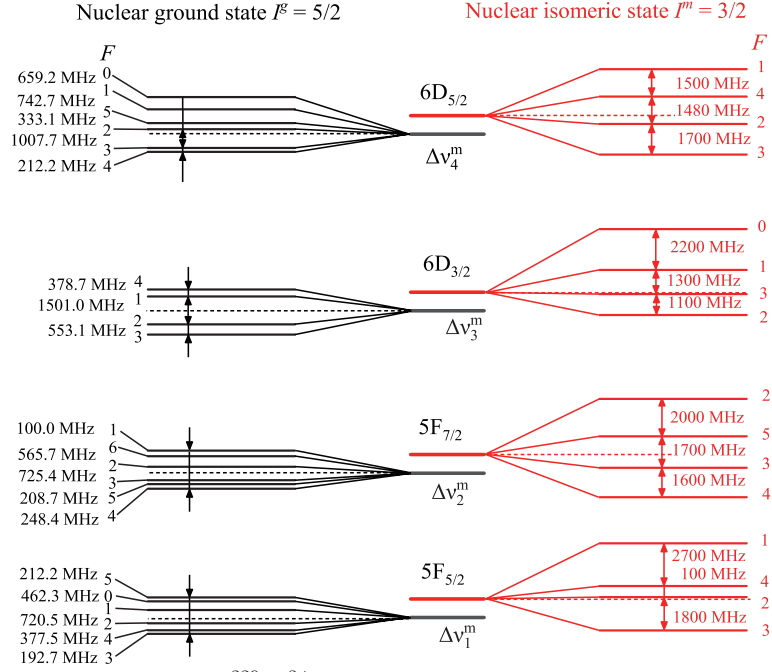}
\caption{\label{AppB B1} The four lowest-lying fine structure levels of
$^{229}$Th$^{3+}$, including hyperfine structures. The integers near atomic levels indicate total angular momentum quantum numbers $F$.
}
\end{figure*}

Therefore, the hyperfine structures of $^{229m}$Th$^{3+}$ can be obtained by
{\small
\begin{eqnarray}
E_{HFS}(J,I^{m},F)=\frac{A^{m}K}{2}+\nonumber\\\frac{B^{m}[\frac{3}{4}K(K+1)-I^{m}(I^{m}+1)J(J+1)]}{2I^{m}(2I^{m}-1)(2J-1)}\label{AppBaq3},
\end{eqnarray}
}
where $K=F(F+1)-J(J+1)-I^{m}(I^{m}+1)$.
 \par 

Based on Eq.~\ref{AppBaq3}, the hyperfine structures of the $^{229m}$Th$^{3+}$ four lowest levels are calculated as shown in Table.~\ref{AppBtab:2}

\begin{table}[htbp]
\footnotesize
\setlength{\tabcolsep}{6mm}
\newcommand{\tabincell}[2]{\begin{tabular}{@{}#1@{}}#2\end{tabular}}
\centering
\caption{ The hyperfine structures of the $^{229m}$Th$^{3+}$ four lowest levels (MHz).}
\begin{tabular}{c c c}
\hline
\hline
\tabincell{c}{Valence orbital} & 
\tabincell{c}{  $F^{m}$  } & 
\tabincell{c}{$^{229m}$Th$^{3+}$} \\
\hline
$5F_{5/2}$ & 1 & 3000(200)\\
&2 & 140(70)\\
&3 & -1700(100)\\
&4 & 270(90)\\
\hline
$5F_{7/2}$ & 2 & 3000(200)\\
&3&-690(70)\\
&4&-2300(100)\\
&5&970(90)\\
\hline
$6D_{3/2}$ & 0& 3400(200)\\
&1&1200(100)\\
&2&-1230(90)\\
&3&-130(90)\\
\hline
$6D_{5/2}$ &1 & 2500(200)
\\
&2 & -450(30)\\
&3 & -2100(100)\\
&4 & 1030(60)\\
\hline
\hline
\end{tabular}
 \label{AppBtab:2}
\end{table} \par

The isomer shift is determined by
\begin{eqnarray}
\Delta\nu^{m}=F^{'}\delta<r^{2}>\label{AppBaq4},
\end{eqnarray}\par

where the difference in the mean-square radii of the isomeric and ground
states in $^{229}$Th is $\delta<r^{2}>$ = 0.012(2)$fm^{2}$ \cite{49} and the field-shift constant $F^{'}$ can be obtained in ref \cite{42}. So the centers of hyperfine structures of $^{229m}$Th$^{3+}$ can be obtained as shown in Table.~\ref{AppBtab:3}.\par

\begin{table}[ht]
\footnotesize
\centering
\caption{ The centers of the hyperfine structures of $^{229}$Th$^{3+}$ four lowest levels (MHz).}
\begin{tabular}{ccc}
\hline
\hline
Valence orbital & $^{229g}$Th$^{3+}$ & 
$^{229m}$Th$^{3+}$ \\
\hline
$5F_{5/2}$ & 0 & 0 \\
$5F_{7/2}$ & 129671430(40) & 129671450(50)
\\
$6D_{3/2}$ & 275596727(31)
 & 275597100(100)\\
$6D_{5/2}$ &434280888(31)
 & 434281300(100)
\\
\hline
\hline
\end{tabular}
 \label{AppBtab:3}
\end{table} 
 \par
Both the hyperfine structures of $^{229m}$Th$^{3+}$ and $^{229g}$Th$^{3+}$ are displayed in Fig.~\ref{AppB B1} and the $\Delta\nu_{i}^{m}$ indicates the isomer shift.  \par

The transition frequency of $\mid5F_{5/2},F=1\rangle$ $\rightarrow$ $\mid6D_{5/2},F=0\rangle$ in $^{229g}$Th$^{3+}$ is 434282480(32) MHz. The hyperfine transitions between the state $5F_{5/2}$ and the state $6D_{5/2}$ of $^{229m}$Th$^{3+}$ is listed in Table.~\ref{AppBtab:4}.\par

\begin{table}[h]
\footnotesize
\centering
\caption{ The hyperfine transitions from the state $5F_{5/2}$ to the state $6D_{5/2}$ of $^{229m}$Th$^{3+}$ (MHz).}
\begin{tabular}{cccc}
\hline
\hline
transition& & transition &  \\
\hline
1 $\rightarrow$ 1 &434280800(300)& 2 $\rightarrow$ 1 & 434283700(200)\\
1 $\rightarrow$ 2 &434277900(200)& 2 $\rightarrow$ 2 & 434280700(100)\\
3 $\rightarrow$ 2& 434282600(100) & 2 $\rightarrow$ 3 & 434279100(200)\\
3 $\rightarrow$ 3& 434280900(200)& & \\
3 $\rightarrow$ 4 &434284000(200) & & \\
\hline
\hline
\end{tabular}
 \label{AppBtab:4}
\end{table} \par
 The transition frequency of $\mid5F_{5/2},F=3\rangle$ $\rightarrow$ $\mid6D_{5/2},F=2\rangle$ in $^{229m}$Th$^{3+}$ is close to the transition frequency of $\mid5F_{5/2},F=1\rangle$ $\rightarrow$ $\mid6D_{5/2},F=0\rangle$ in $^{229g}$Th$^{3+}$.\par
 The transition frequency of $\mid5D_{5/2},F=0\rangle$ $\rightarrow$ $\mid5F_{7/2},F=1\rangle$ in $^{229g}$Th$^{3+}$ is 304610430(34) MHz. The hyperfine transitions between the state $\mid6D_{5/2},F=2\rangle$ and the state $5F_{7/2}$ of $^{229m}$Th$^{3+}$ is listed in Table \ref{AppBtab:5}. As a result, the 690 nm and 984 nm lasers used for cooling of the $^{229g}$Th$^{3+}$ ions can not form the closed optical cycle for laser-cooling the $^{229m}$Th$^{3+}$ ions.  
 \begin{table}[htpb]
 \footnotesize
 \setlength{\tabcolsep}{6mm}
 
\centering
\caption{The hyperfine transitions between the state\\ $\mid6D_{5/2},F=2\rangle$ and the state $5F_{7/2}$ of $^{229m}$Th$^{3+}$ (MHz).}
\begin{tabular}{cc}
\hline
\hline
transition&    \\
\hline
2 $\rightarrow$ 2 & 304606900 (200)\\
2 $\rightarrow$ 3 & 304610500 (100)\\
\hline
\hline
\end{tabular}
 \label{AppBtab:5}
\end{table} \par

\end{document}